# Acoustic Local Positioning with Encoded Emission Beacons

J. Ureña, *Senior, IEEE,* A. Hernández, *Senior, IEEE*, J. J. García, *Member, IEEE,* J. M. Villadangos, M. C. Pérez, *Member, IEEE,* D. Gualda, F. J. Álvarez, *Senior, IEEE,* T. Aguilera

*Abstract*— Acoustic Local Positioning Systems (ALPSs) are an interesting alternative for indoor positioning due to some advantages over other approaches, such as their relatively high accuracy, low cost or room-level signal propagation. Centimeter or fine-grained indoor positioning can be interesting for robot navigation, for guiding a person to a particular art piece in a museum or to a specific product in a shop, for targeted advertising, for augmented reality, etc. In airborne system applications, acoustic positioning can be based on using opportunistic signals or sounds produced by the person or object to be located (e.g. noise from some appliances or the speech from a speaker), or on encoded emission beacons (or anchors) specifically designed for this purpose. This work presents a review of the different challenges that designers of systems based on encoded emission beacons have to face in order to achieve a suitable performance. At the so-called low-level processing, the waveform design (coding and modulation) and the processing of the received signal are key factors to deal with drawbacks such as multipath, multiple access interference, near-far effect, or Doppler shifting. At a high-level processing, the issues to be addressed are related to the distribution of beacons, easy deployment, as well as calibration and positioning algorithms, including possible fusion of information obtained from maps, on-board sensors, etc. Apart from some theoretical discussions, this work also includes the description of an ALPS system actually implemented, installed in a large area and tested for mobile robot navigation. In addition to its practical interest for real applications, airborne ALPSs can be also an excellent platform to test complex algorithms (taking advantage of the low sampling frequency required), which can be adapted later for other positioning systems, such as underwater acoustic systems or UWB RF systems.

*Index Terms*— Acoustic LPS, indoor positioning, ultrasonic signals, beacon deployment and calibration, positioning algorithms.

## I. Introduction

During the last years, there has been an increasing demand for proven technologies that provide services based on the position of people, mobile robots (MR) or other objects across large indoor areas in buildings and surrounding outdoor spaces. For instance, location aware applications, pervasive computing, or augmented reality require positioning data. In contrast to the high degree of implementation in the Global Positioning System (GPS) outdoors, there is no technology consolidated or operating at the same level indoors, yet.

Positioning technologies are usually classified according to the accuracy and the coverage area that can be achieved in applications in which they are used [1]. Effective Local Positioning Systems (LPS) can be divided into five main categories: optical-based, mechanical-based, magnetic-based, acoustic-based and RF-based systems. As a first comparison, Fig. 1 shows the coverage range and accuracy that can be achieved with each technology.

Optical-based systems include those that use vision or infrared sensors. Currently, the use of these systems is booming in dedicated applications with high accuracy, motivated by the enhancement of the sensory performance from cameras (including range measurement in 3D-cameras), the increase in the transmission data rate, the improvement of the computational capacity, and the high degree of development from the image processing algorithms. In [2], there is a classification of this type of systems according to the way in which they obtain the reference information used to carry out the positioning.

The most used mechanical-based systems are those based on Inertial Measurement Units (IMU). These units are composed of 3-axis acceleration and inclination sensors, which allow to obtain successive positions from the variations of velocities and directions. The main problem of such relative positioning systems is that the noise is cumulative and, thus, they must be normally used in combination with other absolute positioning systems [3]. However, as there is no need of any special change in the infrastructure, they can also be used as transition system between zones covered by other positioning technologies.

Magnetic-based systems make use of artificial or natural magnetic field to obtain the position, usually considering the absolute magnetic field value or its variations. With artificial fields, the coverage areas are constrained to the places in which permanent or induced magnets are installed, whereas in the case of using the natural magnetic field the coverage area is global

Manuscript received -, 2017; revised -, 2017; accepted -, 2017.
This work has been supported by the Spanish Ministry of Economy and Competitiveness (TARSIUS project, ref. TIN2015-71564-C4[1-R&4-R], and SOC-PLC project, ref. TEC2015-64835-C3-2-R).

J. Ureña, A. Hernández, J. J. García, J. M. Villadangos, M. C. Pérez and D. Gualda are with the Electronics Department, University of Alcalá, E-28805 Alcalá de Henares (Madrid), Spain (e-mail: jesus.urena@uah.es).

F. J. Álvarez and T. Aguilera are with the Sensory Systems Research Group, University of Extremadura, E-06006 Badajoz, Spain (email: fafranco@unex.es).

[4]. Positioning with this technology has always the problem of disturbances provoked by changes in the environment (furniture, people, …) that affect the magnetic field.

RF-based systems are currently the most used in positioning systems, taking advantage of the installed infrastructure for communication purposes. As is mentioned in [1], any radio signal can be used for indoor positioning at any frequency, signal range and protocol (WIFI, BLE, RFID UWB, LTE, …). Nevertheless, performance levels and applicability greatly vary depending on factors such as the use of pre-existing reference infrastructure, signal ranges, power levels, etc. The main methods used with RF can be based on signal strength fingerprinting (several meters of accuracy with important pre-calibration efforts) or distance-based (where time-of-flight measurements must face harsh environment for indoor signal propagation).

Finally, acoustic-based systems, which we will deal with hereinafter, use sound or ultrasound to estimate the position and, depending on the application, the pose of an object or a person (in general, the target). As can be observed in Fig. 1, this type of technology can reach accuracies around 1cm with coverage distances up to some tens of meters. To measure distances with acoustic signals, the most used methods are Times-of-Flight (ToF) [5], sometimes Time-Differences-of-Flight (TDoF), or phase-coherence [6].

ToF is an absolute method and measures the time that the acoustic signal takes to travel from a transmitter to a receiver. From the ToFs, with the speed of sound, the distances can be computed. On the other hand, phase coherence measures the relative difference of phase between two acoustic sine waves. One of them reaches the receiver from a transmitter at an unknown distance, whereas the other does from a transmitter at a certain reference point. If the difference of distances travelled by both signals is less than one wavelength, the system can obtain the position of the unknown source. Table I summarizes the major advantages and drawbacks of using ToF or phase coherence with acoustic systems [6].

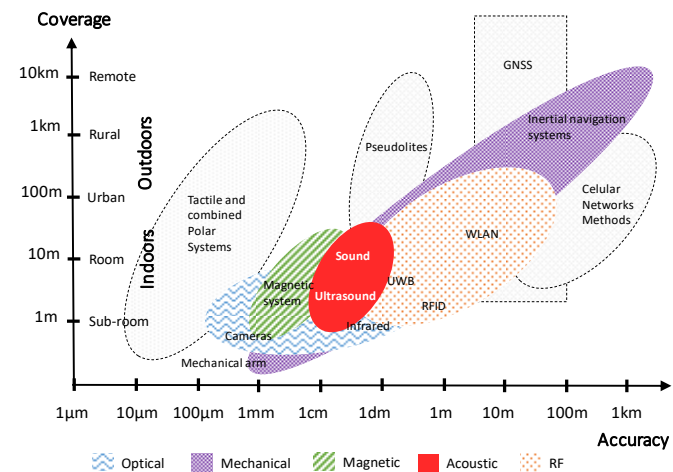

Fig. 1. Accuracy and coverage of acoustic systems regarding other technologies. (Adapted from [1] and [7]).

Independently of the range measurement method used, an ALPS derives the target position after obtaining several distances between the target to be positioned and a set of beacons distributed around the environment at known positions. Two different alternatives can be used: 1) beacons working as emitters and a receiver attached to every target to be positioned (in this case, there is no limitation in the number of receivers and each one can obtain its position independently of the rest keeping then its privacy); and 2) beacons working as receivers and an emitter attached to every target to be positioned. In this case, the number of emitters is limited by the capacity of the transmission channel and the distances are firstly obtained by the infrastructure where the beacons are and, thus, privacy is not guaranteed. In both approaches the channel is shared by different emitter-receiver links, so some kind of medium access control technique must be provided.

As an example, in [8] a room-level accuracy ALPS is presented that covers ranges of more than 10 meters. It consists of a set of transmitters (tags) to be worn by humans or attached to objects, whose positions are required. In this case, a set of receivers is installed on the ceiling or walls of each room. The system uses a Carrier Sense Multiple Access (CSMA) protocol, thus each emitter has to listen for a free channel before attempting to transmit. The low propagation speed of ultrasonic signals implies that the system has to implement a long repetition interval per user and that compromises the maximum throughput achieved (less than 0.5 positions per second and per room).

The system presented in this paper involves another approach for multiuser environments: Direct Sequence Code Division Multiple Access (DS-CDMA). This technique has been used in previous works, in which the encoding of ultrasonic signals can be based on different types of binary sequences: Barker, Golay, Gold, Kasami, $M$-Complementary Sets of Sequences (CSS) or Loosely Synchronized codes[9][10][11].

In ALPS, the absolute distances between the target and the beacons are often obtained (for instance, if the system measures ToFs). The positioning algorithm used in this case is spherical trilateration [12]. It is necessary to provide some kind of synchronization between the beacons and the target. Typically, this synchronization is easily achieved using an additional radio-frequency (RF) link between the tag and the beacons, due to the difference of propagation speeds of electromagnetic and acoustic waves.

TABLE I
ADVANTAGES AND DRAWBACKS OF ACOUSTIC POSITIONING SYSTEMS

| METHOD | ADVANTAGES | DRAWBACKS |
| --- | --- | --- |
| ToF systems | – High accuracy and low latency<br>– Multiple user tracking<br>– Electromagnetic interference immunity<br>– Low cost | – Low update rate<br>– Considerable channel multipath and fading variation over time |
| Coherence Systems | – Low latency<br>– Electromagnetic interference immunity<br>– Multiple user tracking<br>– Low cost | – Time-varying accuracy, (cumulative error over time)<br>– Considerable channel multipath and fading variation over time. |

Although this synchronization can be avoided by using TDoF, it is required an additional beacon. In this case, only the beacons must remain synchronized, and one of them works as the reference beacon. As the system measures differences of distances, the positioning algorithm is based on hyperbolic multilateration that provides a lower accuracy compared to the spherical case [13].

In many cases ALPSs are used as an absolute positioning method for MRs or people; that is, they provide their positions in a global map. MRs use their internal odometer to estimate at a high rate their position and orientation (relative to the initial state), by integrating the number of rotations in their axes. A person can also wear an Inertial Measurement Unit (IMU) with accelerometers and gyroscopes with the same goal, thus obtaining their relative position and orientation. These relative measurements can be very accurate in short integration times, but they also involve cumulative errors. The combination of absolute and relative positioning systems is a good way to have high rates of positioning while cumulative errors are avoided. There are several techniques to merge both positioning systems [14]. For instance, Bayesian methods use statistical distributions to estimate the MR position from a set of measurements [15], dealing with the uncertainty associated to real measurements and with the possibility of adding the a priori knowledge of the positioning system (map information, physical constraints, etc.).

Common approaches are based on the Extended Kalman Filter (EKF), which is optimal for Gaussian noise, as well as on some variants: the Ensemble Kalman Filter (EnKF) that uses a Monte-Carlo method to predict the statistical errors [16]; and the Unscented Kalman Filter (UKF) [17], an estimator for non-Gaussian errors. The work in [18] presents the implementation of an ALPS, which allows a mobile robot to navigate in an extensive area using an H-infinite filter to combine the position measurements provided by the odometer and the ALPS. If the robot is in an area where the ALPS is not available, the H-infinite filter only uses the odometer information and when it reaches an area covered by the ALPS, the absolute position allows to improve the MR position and to cancel the odometer cumulative noise.

Another important aspect for the deployment and use of ALPSs is the search for optimal zones to install the beacons. It implies a tradeoff between coverage area (taking into account a minimum number of beacons at each positioning point), line-of-sights between emitters and receivers, minimization of effects as interferences, multipath, near-far, good geometric configuration (to minimize the so-called Geometric Dilution of Precision, PDOP) and cost (minimum number of beacons). The complexity of this task is solved using meta-heuristic algorithms that provide better results than regular lattices [19].

After the beacons are installed in the environment, it is important to know the precise coordinates for every beacon (with respect to the global coordinate origin). These coordinates are obtained from a calibration process that can be manual, semiautomatic or automatic (also self-calibration). Hand-made calibration processes usually take a long time and require several people carrying out measurements. Self-calibration techniques are very interesting [20] because they allow the system to automatically compute the position of beacons through an inverse positioning algorithm, as the spherical one proposed in [21]. In this work, the positions of the beacons are obtained by taking measurements from several well-known points. Afterwards, the positioning algorithm is applied in an inverse way, i.e. the positions of the beacons are considered as unknown, and the positions from where the distances were obtained -the test points- as known. Also, a self-calibration system for both, the spherical trilateration and the hyperbolic multilateration, can be found in [22]. This system only requires three known test-point positions; the rest of the test points can be unknown positions since they are also computed by the algorithm. Furthermore, the more appropriate region (best zone) to place the unknown test-points is analyzed to achieve an optimal beacon coordinate estimation.

This paper presents a review of different techniques and algorithms, which cover all the signal processing levels, currently used in ALPS. The rest of the paper is organized as follows: Section II deals with the introduction of ALPSs, focused on their use with smart portable devices; Section III describes low-level processing techniques applied to acoustic transmitters; Section IV is devoted to high-level algorithms involved in positioning tasks, calibration and data fusion; Section V explains a practical case of an ALPS implementation; and, finally, conclusions are discussed in Section VI.

## II. ALPS WITH SMART PORTABLE DEVICES

Portable devices, such as smartphones or tablets, offer huge possibilities for the development of Location-Based Services (LBS), what has led to further research efforts on positioning and tracking systems. Outdoor LBS have greatly expanded in recent years thanks to the reliable positioning provided by Global Navigation Satellite Systems (GNSS), which achieves accuracies in the range of a few meters. Unfortunately, these services perform poorly in indoor environments, and there is not an alternative technology mature enough to solve the indoor fine-grained positioning on mobile devices. On the other hand, people spend almost 90% of their time in indoor environments [23]; furthermore, most people in advanced economies own a Smartphone or Tablet, reference [24] reports that 72% of adults in United States, percentage that increases up to 92% if only people aged from 18 to 29 are considered. Similar percentages can be found in several European countries. This explains the great efforts and investments that are currently being made in the development of accurate indoor positioning technologies for mobile devices. If compared with other options based on custom hardware, such as conventional Local Positioning Systems (LPS), the use of smartphones or other mobile devices enhances LBS possibilities, allowing user interaction, augmented reality applications, a wide offer of multimedia services or social networking that can enrich the user experience based on its location. Examples of application include building navigation (in airports, hospitals, factories, malls, …), augmented reality for cultural tourism, gaming, study of point of interests and movement patterns so as to offer targeted advertising or customized services [25].

Existing systems mainly use Radio-Frequency (RF)-based methods, or more recently, acoustic signals, for indoor mobile devices positioning [26][27]. In general, accuracies obtained with RF-based systems are in the range of meters, whereas acoustic systems reach centimeter or even sub-centimeter accuracies [28]. Nevertheless, conventional ultrasonic LPSs (ULPS) require special-purpose speakers, microphones and acquisition hardware (typically operating at around 40 kHz for this type of applications), which are not compatible with current mobile devices.

Recently, there have been proposals that try to merge the advantages of ULPSs with the variety of services and applications that mobile devices can offer. One of the first proposals was the "BeepBeep" system [29] that uses audible acoustic signals, without the aid of external infrastructure apart from the basic hardware included in COTS mobile phones. It is basically a software solution for relative positioning of two mobile devices: each one emits, in turn, a simple linear chirp signal (bandlimited to 2-6kHz), records its own emission and the one from the other device, computes the difference in samples and exchanges the elapsed time with the other device so as to subtract them and obtain the time the sound takes to travel between both devices.

Nevertheless, the obtained accuracies significantly worsen for distances longer than 5m in indoor environments, mainly because of the multipath problem. On the other hand, sound signals can be annoying for users and many applications demand absolute positioning instead of relative one to another mobile device. Another proposal, that overcome some of these drawbacks, is the LOK8 system [25]. It consists of a centralized system, in which the mobile phone to be located transmits a very short designed signal at 21.5kHz; then, a set of four microphones placed at known positions detects the incoming signals from the mobile phone and sends the measured times to a Personal Computer (PC) that runs an asynchronous multilateration algorithm based on TDoF. Results in a 7x7m room show accuracies reaching 10cm. However, centralized systems present some concerns about how the user location information is managed, which becomes more critical when the positioning systems have to be installed in public areas [30]. So, the current trend in this kind of applications is the design of privacy-oriented systems, in which the device to be located is the one that controls and computes its own position, and not a central unit. This is the case of the ALPS system in [26], the winner of the 2015 "Microsoft Indoor Localization Competition-IPSN". It is a privacy-oriented system based on the installation of standard speakers in the environment, which simultaneously transmit a chirp modulated signal with frequency sweeping from 19kHz to 23kHz, just above the human hearing frequency range but still detectable by commercial mobile devices. To allow multiple access transmissions, authors use pulse compression techniques based on Hamming codes, also obtaining tight timing resolution and high robustness to noise. Authors report an accuracy below 10cm in 95% of the cases in a 20mx20m indoor area (this error increases up to 31cm in the IPSN scenario). The ultrasonic data gathered by the mobile phone are sent through a wireless link to a main computer for their processing. A similar approach, but with all the processing tasks performed in the mobile device is presented in [31], for an aided tour navigation application in a museum. In this work, four speakers, controlled by an FPGA central unit board, are placed at known positions of the environment. They simultaneously emit four different Kasami codes every 50ms, which have been previously BPSK modulated at 20kHz. Then, a non-limited number of mobile devices (iOS devices in [31]) capture 75.5ms of the incoming signals at a 48kHz sampling rate. Tests in a laboratory of 5.75mx5.50mx3m show errors that ranges from 3cm, in the best cases, to 90 cm in user positions very affected by multipath or near-far effect. Likewise, in [27] is presented a wireless sensor network infrastructure of synchronized acoustic beacons. The previous systems cope with the maximum sampling rate of commercially available smartphones at 44.1kHz, by emitting signals with frequencies just above the audible human range (typically from 18 to 22kHz). Nevertheless, the available bandwidth to transmit encoded signals is quite narrow, which decreases the signal quality and audible artifacts can appear. The limitation caused by working just above the human hearing range has to be further analyzed in these cases, to avoid potential annoyance to users. A first study with users of different ages has been carried out in [7].

The LOCATE-US system, presented in [32], is a low-cost ALPS for mobile devices in which the incoming signals are not directly captured by the mobile device, but by an attached acquisition module [33][34], thus saving the aforementioned frequency constraints.

Cross-correlation between the transmitted and received signals is commonly used in LPS to maximize the signal-noise ratio (SNR) at the output of the correlator. The development of mobile communications in recent years and the interest of large companies has boosted the proposal of new encoding schemes and modulations [35][36].

The measurement range of a single ALPS is limited. Thus, an application in which the user moves freely in a large indoor environment, will require the use of several ALPSs to cover the entire analysis area. The use of CDMA techniques allows us to assign a set of codes to every ALPS, which identify them, avoiding problems in areas of common coverage. To reduce costs, it is possible to use a particular deployment of different technologies, namely: the acoustic beacons (ALPS) are installed only in critical zones where centimeter positioning is required, such as entrances or exits; whereas in zones where the positioning does not need to be so accurate, the inertial sensors of the mobile phone or even RF signals from the environment can be used to obtain a coarse-grained positioning. The drift errors from the inertial sensors will be corrected when the mobile device reaches one of the ALPS coverage area. Some examples of applications combining several technologies can be found in [25], [31] and [37].

III. LOW-LEVEL PROCESSING IN ALPS

For the sake of clarity, the different algorithms and techniques involved in ALPS has been classified into two different processing level. Fig. 2 summarizes this classification. This Section deals with all the processing techniques used at the last stages before the emissions of the beacons and at the first stages after the signal reception at the receivers.

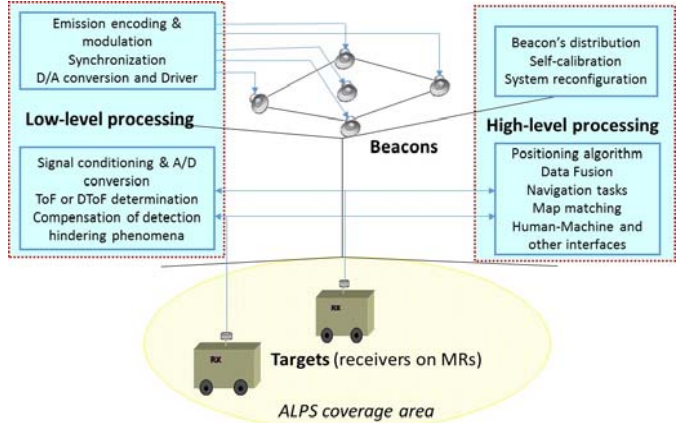

Fig. 2. Example of ALPS with a classification of the algorithms and tasks into low-level and high-level processing.

*A. Narrowband vs Broadband Systems*

First ALPS proposals, dating from the late 1990s and early 2000s, were based on the emission of short and constant-frequency ultrasonic pulses whose arrival was detected by following a simple amplitude or energy thresholding procedure [12][38][39]. This approach significantly reduced the complexity of both the emitter and receiver acoustic modules, but at the expense of providing a limited ranging precision (some tens of cm) with high sensitivity to in-band noise. Furthermore, special attention had to be paid to avoid interference between different emitters, either by making use of time-multiplexing strategies [38] or by developing specific algorithms [39].

To overcome these limitations, broadband signals extensively used in radar systems were soon incorporated in a new generation of ALPS [9][10][40] As it is well know from signal detection theory, the precision of a time-delay measurement is given by [41],

$$\delta t_d = \frac{1}{\beta \cdot \sqrt{2E/N_0}} \quad (1)$$

where $E$ is the signal energy; $N_0$ is the noise power per bandwidth unit; and $\beta$ is the effective or rms (root mean square) bandwidth. Since range and time-delay are directly related through the propagation speed, Eq. (1) indicates that the error in the ranging measurement is inversely proportional to the effective signal bandwidth. However, this bandwidth can be increased by shortening the duration of a continuous frequency pulse, but then the amplitude of the signal should be increased to keep constant the signal energy $E$ appearing in the denominator of Eq. (1). Actually, there is a physical limit for this amplitude imposed by both real transducers and electronics. An alternative is to modulate the original waveform to extend its bandwidth while keeping a constant energy, and use a matched filter in the receiver to detect this signal. This technique is known as pulse compression in radar theory [42] or spread spectrum modulation in communications theory [43]. The use of a matched filter in the demodulation process also ensures the maximization of the output SNR, thus solving the high sensitivity of narrowband systems to the presence of in-band noise. The third shortcoming of these systems identified above, namely, interference between different emissions, can be tackled by choosing an appropriate modulation of the emissions or by a suitable selection of sequences, as detailed next.

*B. CDMA-based Systems*

There have been mainly two alternatives to modulate the emitted waveform with the aim of increasing its effective bandwidth. The first one is based on Linear Frequency Modulation (LFM), where the frequency of a pulsed waveform is linearly increased from $f_1$ to $f_2$ over the duration of the pulse. This is, for example, the strategy followed in [40], where an ALPS based on four transmitters sequentially emitting 1ms-long LFM pulses was proposed. A time-multiplexing strategy was employed in this case to avoid interference between different emissions.

The second option to increase the effective bandwidth is based on Binary Phase Coding, where a long pulse is divided into $N$ subpulses whose phase is selected to be either 0 or $\pi$ radians according to the bits of a certain code. If this code is a pseudo-random (PR) sequence, the waveform approximates a noise-modulated signal with a delta-like auto-correlation function. Fig. 3(a) shows the power spectral density of one of these broadband emissions with $N = 255$ and a pulse frequency of 40 kHz (solid line), together with the spectral density of a typical narrowband emission formed by 20 cycles of a 40 kHz tone (dashed line). Fig. 3(b) shows the delta-like autocorrelation function of the broadband emission. The main advantage of this approach is that different sequences from the same family can be generated with nearly null cross-correlation properties, thus allowing the simultaneous emission of different emitters with very low interference among them. This technique to share the transmission channel among several users by assigning them different modulating PR codes, is known in communications theory as Code Division Multiple Access (CDMA). To date, several CDMA-based ALPS have already been designed that propose the use of different sequences, such as Gold [10], Kasami [9], LS [11] or CSS [44]. In these systems, the receiver's matched filter can be designed as a straightforward digital correlator matched to the modulated waveform [10]. To reduce the hardware implementation complexity, some works propose the cascade combination of two correlators, a first one matched to the subpulse waveform and a second one based on an efficient architecture matched to the binary code [44][45]. These efficient architectures reduce the total number of arithmetic elements required to perform the correlation of an **N**-bit sequence from $O(N)$ to $O(log_2 N)$, thus allowing the actual implementation of a real-time operating system in a hardware platform.

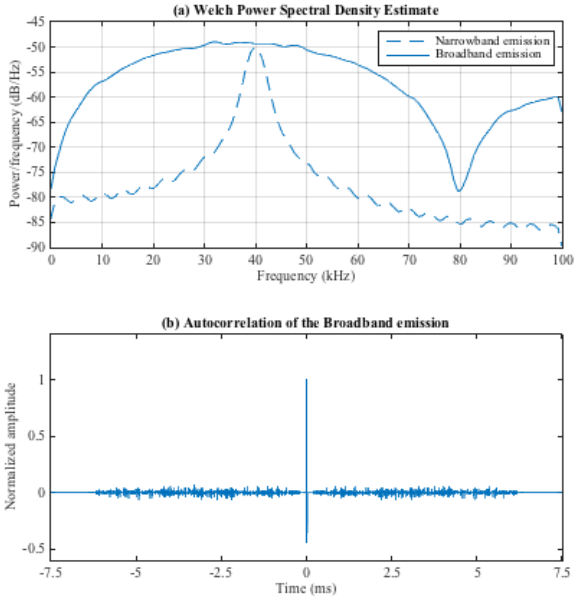

Fig. 3 (a) Power spectral density of a narrowband emission (20 cycles of a 40 kHz tone) and a broadband emission (a 40 kHz pulse BPSK modulated with a 255-bit PR sequence); (b) Autocorrelation of the broadband emission.

In most CDMA-based systems, the time-delay of the received signal is measured when the auto-correlation peak exceeds a certain threshold, improving the precision of the range measurements between one and two orders of magnitude with respect to that of narrowband systems. Also, robustness to in-band noise is significantly improved thanks to the process gain provided by the matched filtering detection, which is proportional to the length of the emitted codes. Unfortunately, the use of simultaneous encoded emissions aggravates the pernicious effect of other phenomena that may hinder signal detection. Some of these phenomena are described below.

*C. Detection Hindering Phenomena*

There are three detection hindering phenomena that have been studied in the context of broadband ALPS, namely, Inter-Symbol and Multiple-Access Interference, Multipath Propagation and Doppler Shift. Although it is out of the scope of this work to present a comprehensive description of the different solutions proposed in the literature, it is worth mentioning the main effect that these phenomena may have on the system performance.

The signal received $r(t)$ in a broadband ALPS with $N$ emitters can be expressed as:

$$r(t) = \sum_{j=1}^{N} A_j \cdot (h_j * g_j)(t - t_j) + n(t) \quad (2)$$

where $g_j(t)$ are the modulated coded signals; $t_j$ and $A_j$ are the ToFs and amplitudes of the signals to be estimated; $n(t)$ represents the noise; and index $j$ runs from 1 to $N$. The effect on the signal of the acoustic channel is introduced in the convolution term $h_j(t)$. This represents the channel impulse response, a priori unknown. The output of the conventional receiver is formed by correlating $r(t)$ with all signal codes. For the $k$-th beacon,

$$R_{rg_k}(t) = A_k \cdot \underbrace{(h_k * R_{g_k g_k})(t - t_k)}_{ISI_k} \\ + \underbrace{\sum_{j \neq k} A_j \cdot (h_j * R_{g_k g_j})(t - t_k)}_{MAI_k} \quad (3) \\ + \eta(t)$$

where $R_{g_k g_j}(t)$ is the cross-correlation of codes $g_k(t)$ and $g_j(t)$. As Eq. (3) outlines, there are two effects which deteriorate the estimation of the ToFs: Inter-Symbol Interference (ISI), which is due to the fact that the limited bandwidth of the acoustic channel lowers the correlation peaks and degrades the signal detection and ranging; and Multiple-Access Interference (MAI) among all the emitted codes in which larger amplitude signals make difficult the detection of weaker signals emitted simultaneously. Combined, both effects can lead to large deviations (outliers) of the ToF estimates from their true values, which can be compensated by using recursive subtractive techniques [46].

MAI might be also avoided using another multiple access technique as Time-Division-Multiple-Access (TDMA), since each emitter has its own slot of time to use the channel. The problem with this technique is the slow positioning rate achieved, mainly due to the low speed of acoustic waves. Nevertheless, an intermediate option between TDMA and CDMA (T-CDMA) has been proposed in [47]. The idea is to insert a certain delay between emissions to have less superposition of signals, but still having a CDMA separation (as some superposition of signals in the channel persists).

On the other hand, the multipath propagation has also been identified as a main cause of degradation in the performance of broadband ALPS. The effect of this phenomenon is critical near room walls and corners, where the strongly reflected signals interfere with the Line-of-Sight (LOS) emissions and deteriorate the ideal correlation properties of these emissions. As a direct consequence of this deterioration, the largest correlation peaks obtained by matched filtering at the receiver do not always correspond to the instant of arrival of the LOS emissions. Recent works propose the use of a Matching Pursuit algorithm to estimate the LOS-TOF in a broadband ALPS and cancel the effect of multipath on signal detection [48][49].

Finally, the low speed of sound in air that allows high resolution positioning is also the cause of a new difficulty in broadband systems. As some authors have pointed out [50][51], the Doppler shift caused by the user's movement in the acoustic encoded signal could make it completely unrecognizable to the receiver. This is a well-known problem in radar theory, whose effects are considerably more pronounced in acoustic systems, since the ratio between the speed of the receiver/reflector and the propagation speed of the signal is typically several orders of magnitude larger than in radar. This problem can be solved by replacing the single correlator for a bank of them, each one matched to different frequency-shifted versions of the code to be detected [52]. Also, some authors have proposed the use of intrinsically Doppler resilient polyphase codes to modulate the emissions of an ALPS [53]. This is, in any case, an interesting and still open field of research.

## IV. HIGH-LEVEL PROCESSING

From a high-level processing point of view, there are several issues that should also be considered when designing an ALPS, namely, distribution of beacons, positioning algorithms, integration of relative positioning systems, calibration algorithms, and the integration of map-matching techniques. This Section deals with the key points for each one.

### A. Distribution of Beacons

The determination of the number of beacons that compose an ALPS and their distribution is one of the first tasks to be carried out, since the cost and the accuracy of the system depend on this distribution. In addition to the coverage area and the cost, one of the most common metric to select the number and distribution of the beacons is the optimization of the measurement uncertainties propagation to the position estimation error [54]. This propagation uncertainty is known as Dilution of Precision (DOP) [55] or Position Dilution of Precision (PDOP) if we consider only variances of coordinates. An estimation of the PDOP can be obtained as:

$$PDOP \approx \frac{\sqrt{\sigma_x^2 + \sigma_y^2 + \sigma_z^2}}{\sigma_m} \quad (4)$$

where $\sigma_x^2, \sigma_y^2, \sigma_z^2$ are the position variances in $X, Y$ and $Z$, respectively, and $\sigma_m$ is the standard deviation in the distance measurements.

Therefore, a suitable geometry for an ALPS structure should provide low PDOP values along the coverage area, thus meaning a system with lower positioning uncertainties.

Another important factor to take into account to determine the structure of an ALPS is the reduced coverage range of the acoustic beacons (few meters). Sometimes, the emitters are placed as close to each other as possible to cover a common positioning space with similar influence of all the beacons (although in this case the PDOP could not be optimal).

The use of metaheuristic optimization methods such as Genetic Algorithms (GA) [56], is one of the most extended procedures to determine the number of beacons and the placement strategy based on the PDOP optimization, considering some restrictions such as the minimum distance among them. GA are proposed in [19][57][58] to solve the sensor placement problem, and, in the particular case of acoustic signals, one of the first studies was reported in [59] and, more recently, in [48].

One of the most compact and portable ALPS, whose first version was released in [60], shows a square structure composed of five beacons (four emitters placed at the vertices and the fifth one at the centre). A similar structure has been recently proposed in [61]. In these cases, the main advantage is that all the beacons have the same orientation and a similar coverage area and, in addition, at the receiver the acoustic powers coming from the different beacons are similar.

### B. Positioning Algorithms

The main objective of a positioning system is to obtain the position of a target inside a coordinate reference. This position is commonly estimated by solving a non-linear system of equations derived from the ranging measurements. There are different techniques depending on the nature of the observations. The most used for acoustic positioning are the ToF and the TDoF.

As mentioned in the Introduction, ToF is based on measuring the time between the signal emission of an active beacon and the reception in the target (that can be a system carrying another receiving beacon or a microphone). This method requires synchronization between the emitter and the receiver, as it is required to know exactly when the emitter starts to emit the signal. After obtaining a set of ToFs, they can be converted into distances through the propagation velocity of sound.

TDoF [62] is the ranging technique used when there is not synchronization between the emitters and the receiver. It is based on the differences of time between the receptions of two or more different emissions, using one of them as a reference. Then the time differences are converted into differences of distances.

When a minimum set of distances or differences of distances is available (at least three for 3D positioning in a hemispace), the position can be obtained by solving the corresponding system of equations (based on either trilateration or multilateration, i.e. either spherical or hyperbolic positioning, respectively). There are two approaches for it: the minimization of a cost function; or directly through geometric methods.

Geometric methods are based on the linearization of the equation system in order to find a closed solution [63]-[65]. The problem of these methods is that they can be very sensible to noise in some positioning areas. Other geometric methods deal with this problem using the Cayley-Menger bideterminant described in [66]. An example of this method for spherical trilateration is shown in [67], and for the hyperbolic multilateration case in [68].

In other approaches, the position is solved by iterating until a minimum is found for a cost function related to the positioning of a target. One of the most used method is Gauss-Newton (GN). Some examples of this strategy can be found in [69] [70]. Note that it is crucial to provide an adequate initialization for this kind of algorithms in order to facilitate its convergence towards a suitable solution. A summary of the mathematical methods for indoor positioning is shown in [71], whereas [68] compares the GN implementation with a geometrical method based on the Cayley-Menger bideterminant. The study shows that the GN method is more accurate (centimeter resolution) than the geometrical one, but with a higher computational cost.

### C. Integration of Relative Positioning

As was previously explained, the position of a target can be updated by using different methods. When the target is moving inside an area covered by one or more ALPSs, not only the information coming from the ALPSs, but also that related to its movement, such as its velocity or its orientation, can be merged.

The movement measurements are usually obtained from onboard sensors: odometers mounted on a Mobile Robot (MR) or Inertial Measurement Units (IMU). This information can be used, together with the ranging measurements and the knowledge of the noise model, to update the position estimate of the target for every iteration. See, for instance, the case in Fig. 4, where a MR, with an odometer onboard, is inside the coverage area of an ALPS.

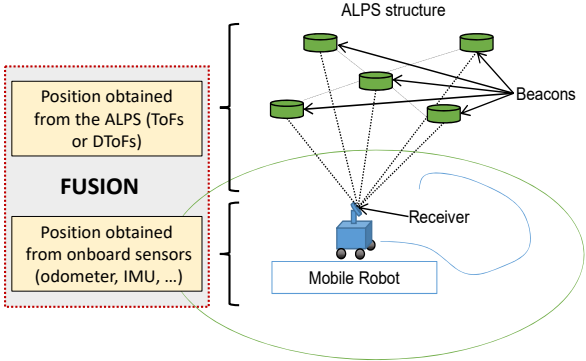

Fig. 4. Diagram of a fusion algorithm when a MR is inside the coverage area of an ALPS.

The position estimation is often represented in the discrete space state as (5)-(6):

$$\mathbf{X}_k = f(\mathbf{X}_{k-1}, \Delta d_{odo}, \Delta \theta_{odo}) + \mathbf{w}_k \qquad (5)$$
$$\mathbf{Z}_k = h(\mathbf{X}_k) + \mathbf{v}_k \qquad (6)$$

where $\mathbf{X}_k$ is the state vector at instant $k$ (usually the target position and its orientation); $f(\mathbf{X}_{k-1}, \Delta d_{odo}, \Delta \theta_{odo})$ represents the relationship between the previous and the current state (that is, $\mathbf{X}_{k-1}$ and $\mathbf{X}_k$) obtained from the onboard sensors and/or a movement model that give the increments of distance and angle ($\Delta d_{odo}, \Delta \theta_{odo}$); $h(\mathbf{X}_k)$ represents the relationship between the state vector $\mathbf{X}_k$ and the ALPS measurements $\mathbf{Z}_k$; and $\mathbf{w}_k$ and $\mathbf{v}_k$ are noise matrices related to the process and measurement equations. The estimated state vector $\widehat{\mathbf{X}}_k$ and its covariance matrix $\widehat{\mathbf{P}}_k$, can be updated by using a filter that combines all the information as the diagram in Fig. 5 shows. The most referenced methods are the Extended Kalman Filter (EKF) [72], the Unscented Kalman Filter (UKF) [17], the H-Infinity filter (H-∞) [74], and the Particle Filter (PF) [75]. Some examples of these methods in ALPSs can be found in [18][76].

*A. Calibration Algorithms*

ALPSs based on ranging measurements need to know the precise positions of the beacons regarding the global coordinate reference. This issue causes a difficulty in terms of portability and adaptability of the systems due to the fact that a manual calibration process requires time and personal efforts. If the acoustic system is composed of a high number of beacons, the calibration can be a hard and tedious task.

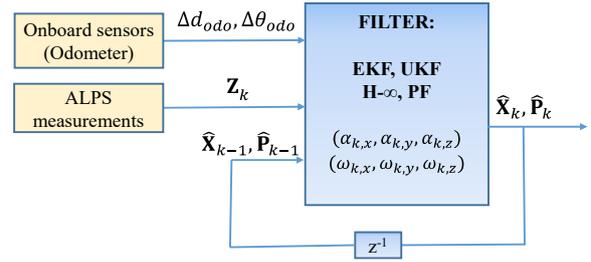

Fig. 5. Diagram of the fusion method for absolute and relative positioning.

One method to facilitate the calibration is the use of inverse positioning to determine the positions of the beacons. This process is based on capturing several measures at known positions in the environment and then obtaining the estimates of the beacon positions by using a positioning algorithm [20][22][78]. The drawback of this method is that it is necessary to accurately know the coordinates of a high number of measurement test points.

Another approach is the use of iterative methods. The work in [21] presents a method, in which the positions of the beacons are estimated by mixing known and unknown test points and applying a Levenberg-Marquardt minimization algorithm. However, it requires a high number of iterations in order to achieve an accurate result. Other proposals use a MR to estimate the position of the beacons while the MR is navigating inside the ALPS coverage area, merging the odometry information with the measured distances [79]. This method was extended to more than one ALPS [80], showing in this case that the error in the beacon position estimates is accumulatively increased and, thus, only a limited number of ULPS can be calibrated in the same process.

*B. Integration of Map-Matching Techniques*

In some occasions, when a target is navigating along indoor environments, some position estimates can be out of the possible geometric boundaries (i.e. the estimate of a target position is behind a wall). Recent solutions in the field of RF positioning [81][82] have incorporated the information from the building map as a constraint. In the case of using a PF, a new stage is added, where the particles that are out of the possible estimates are eliminated. The authors of [83] presents a solution based on a constrained EKF for acoustic positioning that improve the results of the PF approach in terms of computational efficiency.

Another advantage of using mapping constraints is the reduction of the number of sensors needed to cover a positioning area, since the position of the target can be limited to a graph (i.e. a line placed in the middle of a corridor). The work in [84] studies the positioning based on graph information that evaluates the behaviour of an ALPS for different number of beacons.

Additionally, the map information can be used as well when calibrating the position of the beacons in an ALPS. For instance, in [85] a proposal that estimates the position of beacons in a map is presented. In this case, the user inserts several distances from the beacons to the surrounding walls, as well as other heuristic information easily obtained, such as the approximated region where the beacons are (that can be roughly drawn on a map) and the approximated direction of measurement from the beacons to the walls.

## V. Implementation and Applications

From an implementation point of view, previous works have already addressed the challenges coming not only from the positioning issue, but also from the sensory technology used in the ToF or TDoF determination, often presenting similar constraints and issues. In this sense, the authors have already been involved in the real-time implementation of several LPSs based on different technologies, such as radio-frequency [86], infrareds [87], ultrasounds [60], IMU-based approaches [88] or fusion of GPS/RF/ULPS [89].

The design of any acoustic positioning system involves two different modules: the emitters (here beacons) and the receivers. For the sake of clarity, hereinafter, it is considered that the beacons are fixed at certain positions, from where they can transmit the acoustic waves; on the other hand, the receiver is mobile, and often portable. It is worth noting that this approach does not imply any loss of generality, since it could be swapped, thus having a mobile and portable transmitter whereas the beacons would acquire those emissions.

### A. Design of a Beacon Unit

From now on, a single ALPS is the module formed by the acoustic emitters or beacons, as well as by all the electronic equipment necessary to achieve the desired behaviour in the corresponding transmission. The design of this module implies a set of requirements [88], which can be summarized next:
- The first issue to be considered is the selection of the acoustic transducer. Three main aspects should be observed at this point. On one behalf, the acoustic power and the radiation pattern (aperture beam) will define the final coverage area achieved by the ALPS. Furthermore, the transducer response should also be analysed, particularly whether the ultrasonic transmission involves advanced encoding and/or modulation schemes, since they can require specific bandwidths and linearity properties.
- A second issue is the way used by the transducer to access and share the channel. A simultaneous or a multiplexed access protocol have different consequences on the design decisions, which have to be evaluated.
- As has already been mentioned, the bandwidth available in the final beacon unit is a key feature, since it may significantly determine, not only the range of sequences and modulations that can be applied, but also the final performance of the whole system, in terms of precision possible simultaneous emissions, or noise immunity.
- A last detail to be considered is the synchronization issue. This is closely related to the positioning algorithm later implemented. If a spherical positioning algorithm is involved, it is necessary to provide a synchronization link between the beacons and the receiver to achieve a common time frame. On the other hand, hyperbolic algorithms allow this synchronization link to be avoided, although the emitting beacons still need to have the same time reference, in order to properly estimate the TDoFs.

According to the parameters that should be taken into account in the final design, it implies that the electronic system managing the beacons can achieve a significant complexity, mainly due to the encoding and modulation schemes selected in the implementation. Note that the computational load may include correlations, modulations, filtering, or FFTs. Furthermore, the sequences involved in the encoding could be either binary or multi-level, or even complex, thus increasing complexity in real-time hardware architectures.

Fig. 6 shows a general block diagram that can be applied to most ALPS [88]. Four key blocks can be identified: the acoustic transducers, the amplification stage, the AD converters, and, finally, the electronic system in charge of managing the transmission carried out by the beacons available in the ALPS.

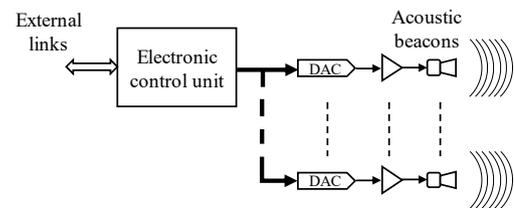

Fig. 6. General block diagram for single ALPS.

With regard to the design of this electronic system, it is possible to distinguish two major trends: one based on general-purpose processors, and another based on FPGA devices [91]. The first one provides a simple and straight way to control beacons in an ALPS, with short development times. Nevertheless, this approach often becomes unsuitable when advanced solutions are proposed in the acoustic signal processing. As an example, it cannot often afford simultaneous access techniques, with synchronous and accurately simultaneous access to several beacons; also, modulations, such as multi-carrier ones, cannot be dealt with. Generally speaking, these proposals achieve a limit when a massive and parallel data processing is necessary. At this point, recent Systems-on-Chip (SoCs), based on FPGAs, allow the designers the integration of the processor's advantages, thanks to the SOC placed in the same die (typically an ARM architecture), with the flexibility and parallelism from the configurable logic in the FPGA device [92] [93].

As a case of study, this last part is dedicated to the description of the LOCATE-US beacon unit, developed by the GEINTRA-US/RF group from the University of Alcalá [94]. In this particular case, the beacon is based on an ultrasonic transducer. This unit was designed keeping in mind that it could be adapted to any possible approach for the signal processing involved in an ALPS. It proposes a flexible architecture where a memory bank is connected to each ultrasonic transducer. This bank memory can be used to store the samples corresponding to any transmission, no matter the type of encoding or the modulation scheme considered. The architecture is based on a FPGA device, consisting of an ARM processor plus some specific advanced peripherals in charge of managing the ultrasonic emissions. Fig. 7 shows its block diagram. It is worth noting that the system can be adapted to any kind of modulation (BPSK, FSK, multi-carrier modulations, etc.) [33], as well as to any kind of sequences (binary, multi-level, complex, etc.) [96].

The system consists of five ultrasonic beacons, placed at the ceiling, in a 70.7x70.7cm square providing a coverage area around 30m$^2$ for a height $h$=3m. Fig. 8 shows two versions of the LOCATE-US beacon unit with different geometrical configurations. The difference is that in the second version the transducer are not coplanar, so the system can also be applied to 3D positioning.

The transducers are hardware synchronized and emit simultaneously or sequentially every certain milliseconds each one (depending on the length of the code and the modulation scheme). A schematic representation of these kinds of emissions can be observed in Fig. 9. In one case, the emissions of all beacons are simultaneous every 20ms (the separation at the receiver is performed by code, CDMA). In the other case, there is a different slot of time for each emission, although some interferences can still be produced due to partial superposition of signals in the channel (the separation at the receiver is carried out by time and by code, T-CDMA). Both schemes permit an emission encoding with a 255-bit Kasami sequence, BPSK-modulated with a symbol composed of two cycles of a sine carrier at a frequency fc=41.67kHz. The sampling frequency is fs=500kHz, thus giving an oversampling of fs/fc=12.

After each period of emissions, a non-limited number of portable receivers (smartphones, tablets or similar) can compute their position autonomously by hyperbolic trilateration. As an advantage, this system does not require synchronization between the beacons and the receivers. If spherical trilateration is required or multiple beacon units need to be synchronized to provide greater coverage, a RF module synchronism indicates the instant of transmission to the receivers and/or to the rest of slaves ALPS units.

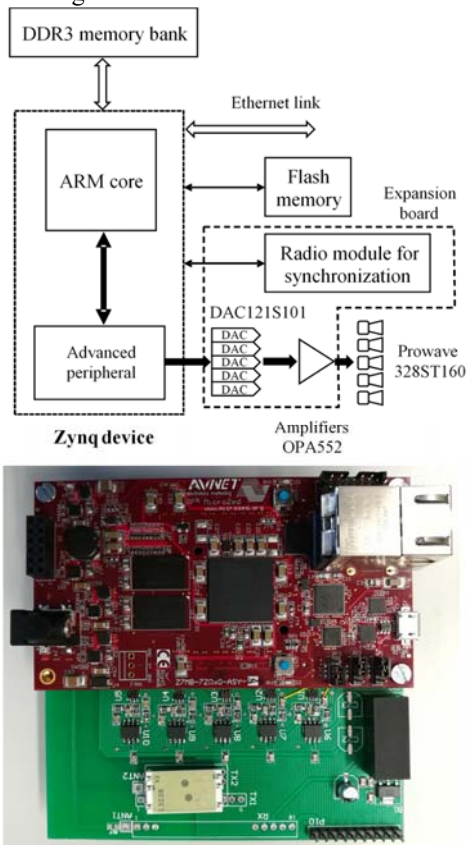

Fig. 7. Block diagram (up) and general view (down) of the ALPS unit designed to control five acoustic beacons, based on a Microzed board, for LOCATE-US, developed by the GEINTRA-US/RF group, University of Alcala [94].

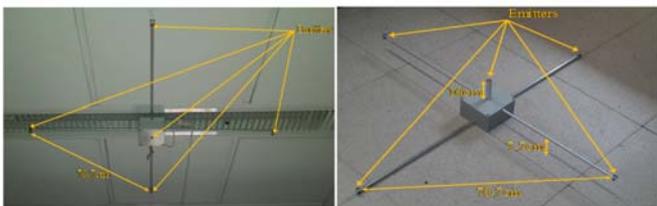

Fig. 8. View of the LOCATE-US beacon unit: version for 2D positioning (left); version more suitable for 3D positioning (right) since the transducers are not coplanar.

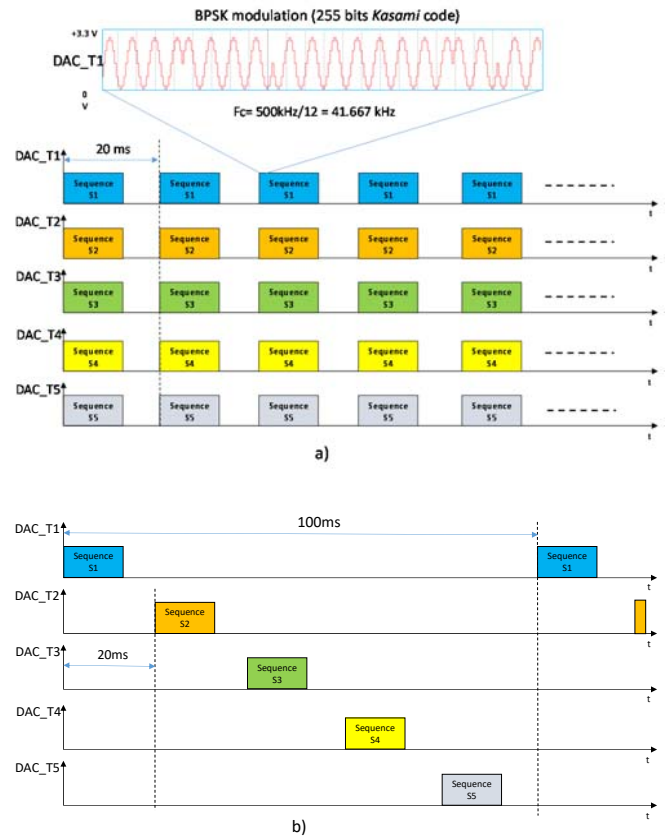

Fig. 9. Two schemes of beacon transmissions: a) CDMA with simultaneous emissions every 20ms; b) T-CDMA with a certain separation in the emissions (period of 100ms).

The 328ST160 transducers used have enough bandwidth for different types of acoustic signal modulations [95]. The transducer frequency response is defined by a flat region between two resonant frequencies (34kHz and 47kHz), so the modulation carrier frequency has been approximately centred between both, thus obtaining a bandwidth of about 12kHz and a nearly constant phase response.

### B. Design of a Receiver

One important aspect to be remarked from the very beginning is that the design of a receiver is not directly related to the design of the beacons in an ALPS. It is clear that some aspects have to be common in both cases, such as frequencies of interest, bandwidths, or sequences employed. Nevertheless, after these general issues, the receiver can follow its own development, as far as it meets certain compliance requirements.

Fig. 10 depicts a general scheme with the basic blocks included in a possible receiver. The ultrasonic transducer should be selected keeping in mind three features: suitable frequency response and compatibility with the beacons one; enough bandwidth to recover the transmitted signal; and enough sensitivity and aperture beam to ensure a certain coverage area. Concerning the pre-amplification stage, it has to be designed with the same considerations, trying to avoid any saturation or non-linear effect coming from non-desired frequency components, out of the range of interest, thus sometimes implying filtering.

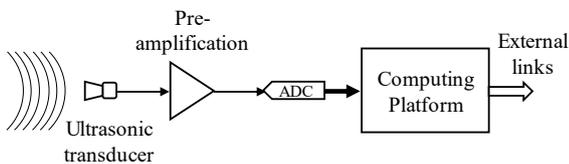

Fig. 10. General scheme in the design of a receiver for an ALPS.

Finally, after acquiring the pre-amplified signal, the computing platform should be capable of processing the ultrasonic transmission, demodulating the signal and then correlating it with the emitted sequences to determine the ToFs or the TDoFs, whether there is or not a synchronization link with beacons. This is likely the most computationally expensive point of the design, since the complexity of operations can easily increase, depending on the selected schemes and sequences, whereas some real-time requirements can arise according to certain applications. The need of having available these ToFs/TDoFs at a certain rate has an influence on the type of architecture that can be proposed for the implementation of the receiver signal processing. Again, most restrictive applications are prone to require parallel approaches, such as those based on FPGA devices. Nevertheless, if timing constraints are not so demanding, processor-based solutions can also offer an alternative, often cheaper and with a shorter development time, for the receiver design.

For the LOCATE-US ALPS as a study case, the receiver architecture has been based on low-cost Cortex M3 microcontroller STM32F103 (see Fig. 11) [11]. It includes a MEMS microphone, a high-bandwidth amplifier and an internally configurable high-pass filter (SPU0414HR5H-S). A programmable gain amplifier allows to dynamically adjust the level of the received signal at the input of the ADC.

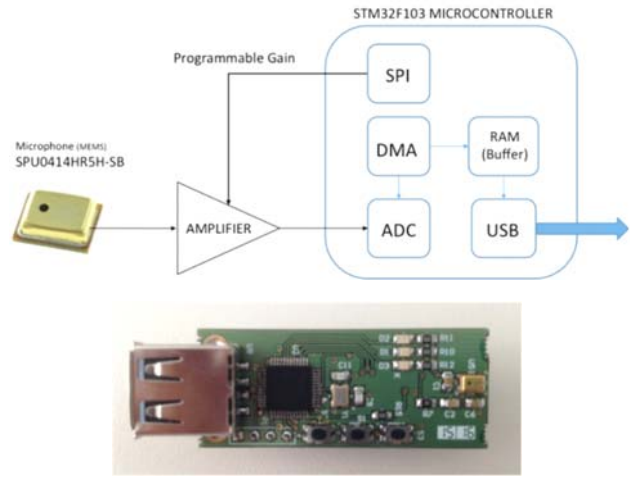

Fig. 11. Block diagram (up) and general view (down) a receiver based in Cortex-M3 microcontroller for LOCATE-US, developed by the GEINTRA-US/RF group, University of Alcala [11].

An important feature in the receiver module is the size of the acquisition buffer, constrained by the size of the internal memory available in the microcontroller. This parameter, together with the sampling frequency, influences on the acquisition time, necessary to obtain a position. In order to shorten this time, it is possible to reduce the emission time for the sequences, or the multiplexing time depending on the transmission scheme used (see Fig. 9), as well as the sampling frequency by downsampling by a factor $N$, even if resolution can be degraded in the ToF or TDoF determination. This affects the positioning error of the receiver, particularly in the case of hyperbolic multilateration (when synchronism between emitters and receiver is avoided). Note that a sampling frequency of 100kHz, with a downsampling factor of 5, provides suitable results, as will be described later. The necessary buffer in the CDMA scheme with 255-bits Kasami sequences (2 cycles per symbol) for an emission period of 20ms should be at least 2000 samples long, in order to guarantee at least one sampling period. On the other hand, in T-CDMA this value is multiplied by the number of transducers, if the multiplexing time is equal to the emission period in CDMA. The main advantage provided by the T-CDMA scheme is that the MAI effect is discarded; furthermore, the algorithm of the DToFs determination is less complex since the transducer emissions are always sorted after performing the correlation of the received signal with the corresponding codes or patterns assigned.

Fig. 12 and Fig. 13 show the results for the signal received from the five beacons (B1-B5) and the correlations obtained in the CDMA and T-CDMA schemes, respectively. When obtaining these results, the receiver has been placed below the central beacon at a distance of 3m from it. It is possible to observe that the overlapping effect in CDMA, coming from the simultaneous emissions of the transducers, can degrade the peaks in the correlation functions due to the MAI effect.

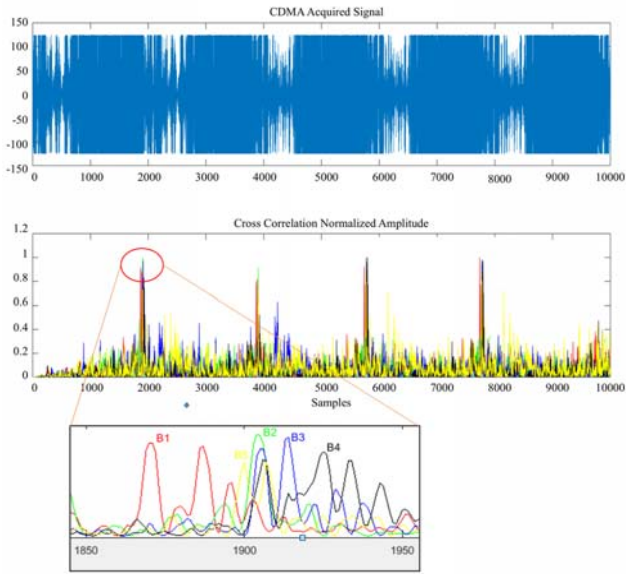

Fig. 12. CDMA scheme: received signal (top); correlated signal for each beacon (B1-B5) (middle); and detail of the correlated signals (bottom).

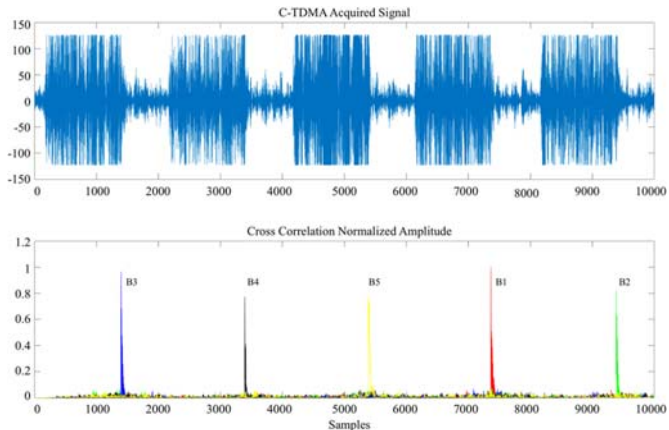

Fig. 13 T-CDMA scheme: received signal (top) and correlated signal for each beacon (B1-B5) (bottom).

With regard to the experimental results, the LOCATE-US ALPS, has allowed a complete set of tests to be obtained in order to verify the different mentioned approaches, not only in the low-level but also in the high-level processing. As an example, for the configuration presented in Fig. 14 (with a single ALPS installed in the ceiling), a microphone has been placed at a grid of points on the floor. The emitted signals have been encoded with a 255-bit Kasami sequence, BPSK modulated with two periods of a sine carrier at 41.67kHz, which has been sampled at 500kHz in the receiver. Afterwards, the position is estimated with a Gauss Newton hyperbolic multilateration algorithm.

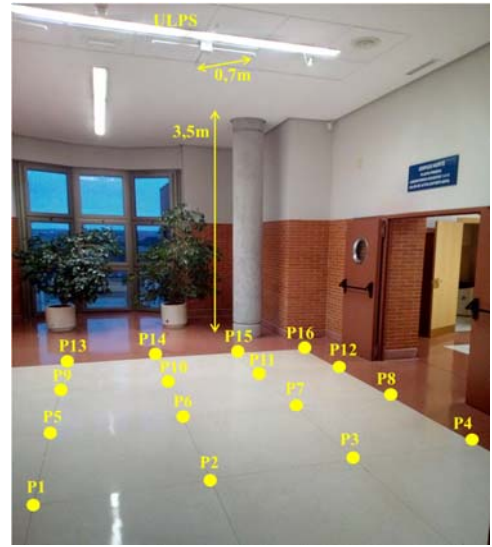

Fig. 14. Configuration of a single ALPS (LOCATE-US) installed on the ceiling and a grid of points to take measurements on the floor.

Fig. 15 shows the cloud of points obtained after 100 trials at each point. Note that the errors depend on the position of the microphone due to the different PDOP and to the different propagation paths followed by the acoustic waves.

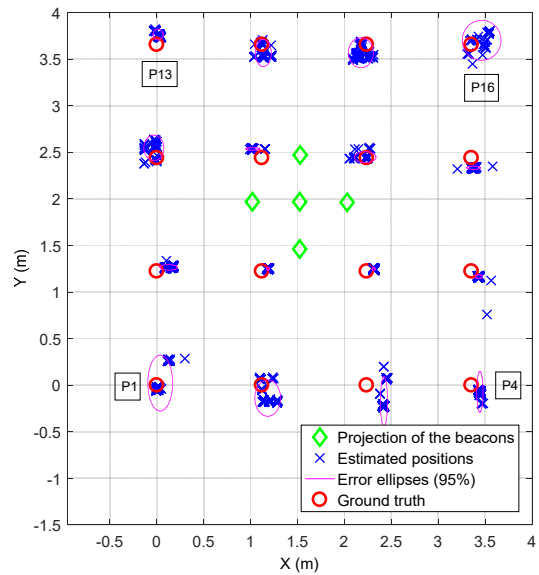

Fig. 15. Cloud of points around the ground truth and error ellipses for 100 trials at each test point.

A quantization of the error for these measurements can be derived from the CDF representations plotted in Fig. 16. For the most centred test points (P6, P7, P10 and P11) the error is always below 15cm. Considering all the points the error is bellow 20cm for 90% of the cases and never higher than 30 cm. It is worth noting that these results have been obtained for an ALPS with beacons very close one each other, using TDoF and with a relatively high common coverage area.

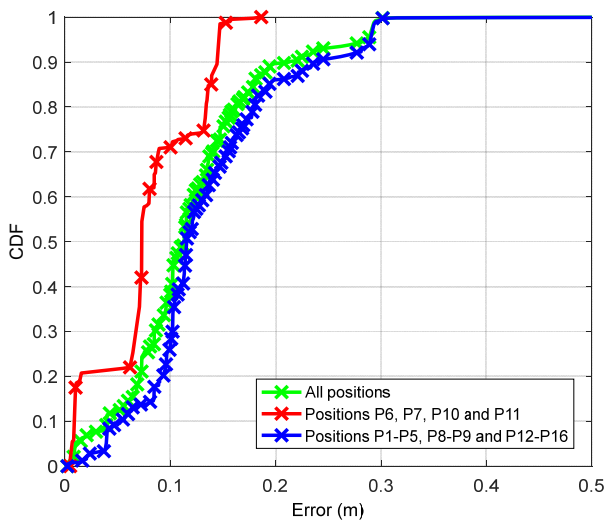

Fig. 16. Cumulative Distribution Functions (CDF) for the trials shown in Fig. 15.

Other tests have been performed with similar ULPSs. In [97] the influence of the sequence encoding involved in the acoustic transmission is analysed for different types of sequences (Kasami and LS codes), and how they can be helpful in order to mitigate some adverse effects, such as multipath or near-far effect. The low-level processing is further studied in [98][99], where the positioning results in adverse environmental conditions can be improved by implementing a generalized cross-correlation together with a PATH filter, instead of a classical matched filter version. This approach results in a significantly better performance for multipath environments.

On the other hand, concerning the high-level processing, in [100] an experimental scenario is described, where multiple LOCATE-US ALPSs have been deployed in a large environment to provide a non-continuous coverage area. The proposal applies an H-infinite filter to merge information, not only from ALPS measurements, but also from the odometer system on board the robot. This test shows the feasibility of covering extensive areas with multiple ALPSs, by choosing a suitable beacon distribution and an appropriate merging method for the positioning algorithm in the high-level processing. Finally, an example of a simultaneous calibration and navigation process for a large environment with multiple ALPSs can be observed in [101]. Note that the beacon positions are not known a priori, and the robot employs the first positions in its navigation (where errors from the odometer are still reduced) to estimate the beacon positions. Afterwards, these estimates for the beacons are used in a common positioning algorithm to follow the robot trajectory.

## VI. CONCLUSIONS

This work has described the different challenges that the design of an acoustic local positioning system must face in order to achieve a suitable performance. The description of this design process has been focused on both the low-level and a high-level signal processing. In the first case, the waveform conformation (coding and modulation) and the processing of the received signal involve significant issues in order to deal with drawbacks such as multipath, multiple access interference, near-far effect, or Doppler shifting. On the other hand, the high-level processing is often related to the distribution of beacons, easy deployment, as well as calibration and positioning algorithms, including possible fusion of information obtained from maps, on-board sensors, etc. In both cases, this work provides a complete review of previous works, as well as some theoretical discussions with regard to the mentioned topics. Furthermore, the description of the LOCATE-US ALPS system, developed by the GEINTRA-US/RF group (University of Alcalá), has been briefly included, together with experimental results in extended coverage areas and tests for mobile robot navigation.

REFERENCES

[1] R. Mautz, "Indoor Positioning Technologies" *[Habilitation Thesis]*, Institute of Geodesy and Photogrammetry, Department of Civil, Environmental and Geomatic Engineering, ETH Zurich, 2012.
[2] R. Mautz and S. Tilch. "Survey of optical indoor positioning systems", in *Indoor Positioning and Indoor Navigation (IPIN), 2011 International Conference on*, pp 1-7, 2011.
[3] A.R. Jiménez, F.S. Granja, J.C. Honorato, and J.I. Rosas. "Pedestrian indoor navigation by aiding a foot-mounted IMU with RFID signal strength measurements", in *Indoor Positioning and Indoor Navigation (IPIN), 2010 International Conference on*, pp 1-7, Sept 2010.
[4] Janne Haverinen and Anssi Kemppainen. "Global indoor self-localization based on the ambient magnetic field", *Robotics and Autonomous Systems,* 57(10): pp 1028-1035, 2009.
[5] A. R. Jiménez and F. Seco, "Precise localisation of archaeological findings with a new ultrasonic 3D positioning sensor", *Sensors and Actuators A: Physical*, vol. 123-124, pp. 224-233, 2005.
[6] Y. Cheng, X. Wang, T. Caelli, and B. Moran, "Target tracking and localization with ambiguous phase measurements of sensor networks", in *2011 IEEE International Conference on Acoustics, Speech and Signal Processing (ICASSP)*, pp. 4048-4051, 2011.
[7] Sergio I. Lopes, *In search of reliable centimeter-level indoor positioning. A smartphone-based approach*, PhD Thesis. University of Aveiro, 2014 (last access realized on June 2017). http://ria.ua.pt/bitstream/10773/14100/1/in%20search%20of%20reliable%20centimeter-level%20indoor%20positioning.pdf
[8] S. Holm, "Airborne Ultrasound Data Communications: The Core of an Indoor Positioning System", in *Proc. IEEE Ultrasonics Symposium*, pp. 1801-1804, 2005.
[9] Jesús Ureña, Manuel Mazo, J. Jesús García, Álvaro Hernández, Emilio Bueno. "Correlation Detector Based on a FPGA for Ultrasonic Sensors", *Microprocessor and Microsystems*, vol. 23, no. 1, pp. 25-33. 1999.
[10] M. Hazas and A. Ward, "A novel broadband ultrasonic location system", in *Proc. Of UbiComp 2002: Ubiquitous Computing*, pp. 264-280, 2002.
[11] M. C. Pérez, J. Ureña, A. Hernández, et al., "Ultrasonic beacon-based Local Positioning System using Loosely Synchronous codes", in *Proc. IEEE International Symposium on Intelligent Signal Processing (WISP 2007)*, pp. 1-6, 2007.
[12] A. Ward, A. Jones, and A. Hopper. "A new location technique for the active office", *IEEE Personal Communications*, vol. 4(5), pp. 42–47, 1997.
[13] M. Deffenbaugh, J. G. Bellingham, and H. Schmidt, "The relationship between spherical and hyperbolic positioning", in *Proc. MTS/IEEE Prospects for the 21st Century OCEANS '96*, vol. 2, pp. 590-595, 1996.
[14] L. Kleeman, "Advanced sonar and odometry error modeling for simultaneous localization and map building", in *Proc. of the Intelligent Robots and Systems (IROS'03)*, vol. 1, pp. 699-704, 2003.
[15] D. Fox, J. Hightower, L. Liao, D. Schulz and G. Borrello, "Bayesian filters for location estimation", *IEEE Pervasive Computing*, vol. 2, no. 3, pp. 24-33, 2003.
[16] G. Evensen, "The ensemble Kalman filter: Theoretical formulation and practical implementation", *Ocean Dynamics*, vol. 53, pp. 343-367, 2003.

[17] E. A. Wan and R. van der Merwe, "The unscented Kalman filter for nonlinear estimation", in *IEEE 2000 Adaptive Systems for Signal Processing, Communications, and Control Symposium (AS-SPCC 2000)*, pp.153-158, 2000.

[18] D. Ruiz, E. García, J. Ureña, D. de Diego, D. Gualda and J. C. García, "Extensive Ultrasonic Local Positioning System for navigating with mobile robots", in *2013 10th Workshop on Positioning Navigation and Communication (WPNC)*, pp. 1-6, 2013.

[19] J. O. Roa, A. R. Jiménez, F. Seco, J. C. Prieto and J. L. Ealo, "Optimal Placement of Sensors for Trilateration: Regular Lattices vs Meta-heuristic Solutions", in *Proc. of the 11th International Conference on Computer Aided Systems Theory (EUROCAST'07)*, pp. 780-787, 2007.

[20] A. Mahajan and F. Figueroa, "An Automatic Self Installation and Calibration Method for a 3D Position Sensing System using Ultrasonics", *Robotics and Autonomous Systems*, vol. 28, no. 4, pp. 281-294, 1999.

[21] P. Duff, and H. Muller. "Autocalibration Algorithm for Ultrasonic Location Systems", in *Proc. of the Seventh IEEE International Symposium on Wearable Computers*, 2003.

[22] F. Daniel Ruiz, J. Ureña, J. M. Villadangos, I. Gude, J. J. García, A. Hernández and A. Jiménez, "Optimal Test-Point Positions for Calibrating an Ultrasonic LPS System", in *Proc. of 13th IEEE International Conference on Emerging Technologies and Factory Automation (ETFA)*, 2009.

[23] L. K. Wang and M. H. S. Wang, *Handbook of Environmental Engineering*, Springer, 2016.

[24] J. Pushter, *Smartphone Ownership and Internet Usage Continues to Climb in Emerging Economies*, Pew Research Center, 2016.

[25] V. Filonenko, C. Cullen and J. Carswell, "Indoor Positioning for Smartphones using asynchronous ultrasound trilateration", *ISPRS International Journal of Geo-Information*, 2013.

[26] P. Lazik and A. Rowe, "Indoor Pseudo-ranging of Mobile Devices using Ultrasonic Chirps", in *Proc. of the 10th ACM Conference on Embedded Network Sensor Systems*, pp. 99-112, 2012.

[27] S. I. Lopes, J. M. N. Vieira, J. Reis and D. Albuquerque, "Accurate Smartphone indoor positioning using WSN infrastructure and non-invasive audio for TDoA estimation", *Pervasive and Mobile Computing*, pp. 1-18, 2014.

[28] J. C. Prieto, A. R. Jiménez, J. Guevara, J. L. Ealo, F. Seco, J. O. Roa and F. Ramos, "Performance evaluation of 3D-LOCUS advanced acoustic LPS", *IEEE Trans. Instrum. Measurements*, vol. 58, no. 8, pp. 2385-2395, 2009.

[29] C. Peng, G. Shen, Y. Zhang, Y. Li and K. Tan, "BeepBeep: A High Accuracy Acoustic Ranging System using COTS Mobile Devices", in *Proc. of the 5th International Conference on Embedded Networked Sensor Systems (ACM Sensys)*, pp. 1-14, 2007.

[30] M. Langheinrich, "Privacy by design- principles of privacy aware ubiquitous systems". In *Proc. of UbiComp 2001: Ubiquitous Computing*, pp. 273-291, 2001.

[31] T. Aguilera, J. A. Paredes, F. J. Álvarez, J. I. Suárez and A. Hernández, "Acoustic Local Positioning System Using an iOS device", in *2013 International Conference on Indoor Positioning and Indoor Navigation*, pp. 1-5, 2013.

[32] E. Díaz, M. C. Pérez, D. Gualda, J. M. Villadangos, J. Ureña and J. J. García, "Ultrasonic indoor positioning for smart environments: a mobile application", in *Proc. of The Experiment International Conference (EXP.AT'17)*, 2017.

[33] A. Lindo, E. García, J. Ureña, M. C. Pérez, A. Hernández, "Multiband Waveform Design for an Ultrasonic Indoor Positioning System", *IEEE Sensors Journal*, vol. 15, no. 12, 2015.

[34] M. C. Pérez, D. Gualda, J. M. Villadangos, J. Ureña, P. Pajuelo, E. Díaz and E. García, "Android application for indoor positioning of mobile devices using ultrasonic signals", in *Proc. of 2016 International Conference on Indoor Positioning and Indoor Navigation (IPIN)*, 2016.

[35] D. F. Albuquerque, J. M. Vieira, I. Lopes, C. A. C. Bastos and P. J. S. G. Ferreira, "OFDM Pulse design with low PAPR for Ultrasonic Location Positioning Systems", in *Proc. of 2013 International Conference on Indoor Positioning and Indoor Navigation*, pp. 34-35, 2013.

[36] E. García, J. Ureña, A. Hernández and F. Nombela, "Discrete multitone modulation for Ultrasonic Indoor Positioning Systems", in *Proc. of 2015 International Conference on Indoor Positioning and Indoor Navigation*, pp. 1-4, 2015.

[37] Wonho Kang, Seongho Nam, Youngnam Han, and Sookjin Lee, "Improved heading estimation for smartphone-based indoor positioning systems", in *2012 IEEE 23rd Intern. Symposium on Personal Indoor and Mobile Radio Communications (PIMRC)*, pp. 2449-2453, 2012.

[38] E. Foxlin, M. Harrington and G. Pfeifer, "Constellation™: A Wide - Range Wireless Motion-Tracking System for Augmented Reality and Virtual Set Applications", in *SIGGRAPH*, 1998.

[39] N. B. Priyantha, A. Chakraborty and H. Balakrishnan, "The Cricket location-support system", in *Proc. of the 6th annual international conference on Mobile computing and networking (MobiCom'00)*, pp. 32-43. 2000.

[40] M. R. McCarthy and H. L. Muller, "RF free ultrasonic positioning", in *Seventh IEEE International Symposium on Wearable Computers Proc.*, pp. 79-85, 2003.

[41] S. M. Kay, *Fundamentals of Statistical Signal Processing: Estimation Theory*, Prentice Hall Signal Processing Series, 1993.

[42] M. I. Skolnik, *Introduction to Radar Systems*, McGraw-Hill, 2001.

[43] J. G. Proakis and M. Salehi, *Communication Systems Engineering*, Prentice Hall, 2nd Ed., 2002.

[44] M. C. Pérez, R. Sanz, J. Ureña, Á. Hernández, C. De Marziani and F. J. Álvarez, "Correlator Implementation for Orthogonal CSS Used in an Ultrasonic LPS", *IEEE Sensors Journal*, vol. 12, no. 9, pp. 2807-2816, 2012.

[45] M. C. Pérez, J. Ureña, C. de Marziani, Á. Hernández, J. J. García and F. J. Álvarez, "Very efficient correlator for loosely synchronised codes," *Electronics Letters*, vol. 46, no. 16, pp. 1127-1129, 2010.

[46] F. Seco, J. C. Prieto, A. J. Ruiz and J. Guevara, "Compensation of Multiple Access Interference Effects in CDMA-Based Acoustic Positioning Systems", *IEEE Trans. on Instrumentation and Measurement*, 2014.

[47] J.M. Villadangos, J. Ureña, J. Jesús García, M. Mazo, A. Hernández, A. Jiménez, D. Ruíz and C. De Marziani, "Measuring Time-of-Flight in an Ultrasonic LPS System using Generalized Cross-Correlation", *Sensors*, vol. 11, pp. 10326-10342, 2011.

[48] F. J. Álvarez, T. Aguilera and R. López-Valcarce, "CDMA-based acoustic local positioning system for portable devices with multipath cancellation", *Digital Signal Processing*, vol. 62, pp. 38-51, 2017.

[49] T. Aguilera, F. J. Álvarez, D. Gualda, J. M. Villadangos, A. Hernández and J. Ureña, "Performance improvement of an Ultrasonic LPS by applying a Multipath Compensation Algorithm", in *2017 IEEE International Instrumentation and Measurement Technology Conference (I2MTC)*, 2017.

[50] M. Alloulah and M. Hazas, "An efficient CDMA core for indoor acoustic position sensing", in *Proc. of the 2010 International Conference on Indoor Positioning and Indoor Navigation (IPIN)*, 2010.

[51] J.A. Paredes, T. Aguilera, F.J. Álvarez, J. Lozano and J. Morera, "Analysis of Doppler effect on the pulse compression of different codes emitted by an ultrasonic LPS", *Sensors*, vol. 11, pp. 10765-10784, 2011.

[52] F. J. Álvarez, A. Hernández, J. A. Moreno, M. C. Pérez, J. Ureña and C. De Marziani, "Doppler-tolerant receiver for an ultrasonic LPS based on Kasami sequences", *Sensors and Actuators A*, vol. 189, pp. 238-253, 2013.

[53] J. A. Paredes, T. Aguilera, F. J. Álvarez, J. A. Fernández and J. Morera, "New pseudo-orthogonal family of polyphase codes to improve Doppler resilience", in *Proc. of 2013 International Conference on Indoor Positioning and Indoor Navigation*, 2013.

[54] R. Kaune, "Accuracy studies for TDOA and TOA localization", in *Proc. of International Conference on Information Fusion (FUSION 2012)*, pp. 408-415, 2012.

[55] R. B. Langley, "Dilution of precision", *GPS World*, vol. 10, no. 5, pp. 52-59, 1999.

[56] J. H. Holland, *Adaption in Natural and Artificial Systems*, Ann Arbor: The University of Michigan Press, 1975.

[57] M. Laguna, J. O. Roa, A. R. Jiménez and F. Seco, "Diversified local search for the optimal layout of beacons in an indoor positioning system", *IIE Transactions*, vol. 41, no. 3, pp. 247–259, 2009.

[58] T. Leune, T. Wehs, M. Janssen, C. Koch and G. von Colln, "Optimization of wireless locating in complex environments by placement of anchor nodes with evolutionary algorithms", in *Proc. of 2013 IEEE 18th Conference on Emerging Technologies & Factory Automation (ETFA)*, 2013.

[59] P. K. Ray and A. Mahajan, "A genetic algorithm-based approach to calculate the optimal configuration of ultrasonic sensors in a 3D position estimation system", *Robotics and Autonomous Systems*, vol. 41, no. 4, pp. 165-177, 2002.

[60] J. Ureña, A. Hernández, A. Jiménez, J. M. Villadangos, M. Mazo, J. C. García, J. J. García, F. J. Álvarez, C. De Marziani, M. C. Pérez, J. A. Jiménez, A. R. Jiménez and F. Seco, "Advanced sensorial system for an acoustic LPS", *Microprocessors and Microsystems*, vol. 31, no. 6, pp. 393-401, 2007.

[61] J. Li, G. Han, C. Zhu and G. Sun, "An Indoor Ultrasonic Positioning System Based on TOA for Internet of Things", *Mobile Information Systems*, vol. 2016, pp. 1-10, 2016.

[62] J. Delosme, M. Morf and B. Friedlander, "Source location from time differences of arrival: Identifiability and estimation", in *Proc. of IEEE International Conference on Acoustics, Speech, and Signal Processing (ICASSP '80)*, 1980.

[63] W. Navidi, W. S. Murphy and W. Hereman, "Statistical methods in surveying by trilateration", *Computational Statistics & Data Analysis*, vol. 27, no. 2, pp. 209-227, 1998.

[64] J. J. Caffery, "A new approach to the geometry of TOA location", in *52nd Vehicular Technology Conference Fall 2000 (IEEE VTS Fall VTC2000)*, 2000.

[65] S. C. Nardone and M. L. Graham, "A closed-form solution to bearings-only target motion analysis", *IEEE Journal of Oceanic Engineering*, vol. 22, no. 1, pp. 168-178, 1997.

[66] D. M. Y. Sommerville, "An Introduction to the Geometry of n Dimensions", Methuen & Co, London, vol. 2, 1929.

[67] F. Thomas and L. Ros, "Revisiting trilateration for robot localization", *IEEE Transactions on Robotics*, vol. 21, no. 1, pp. 93-101, 2005.

[68] D. Ruiz, J. Ureña, I. Gude, J. M. Villadangos, J. C. Garcia, C. Perez and E. Garcia, "New iterative algorithm for hyperbolic positioning used in an Ultrasonic Local Positioning System", in *2009 IEEE Conference on Emerging Technologies & Factory Automation (ETFA)*, 2009.

[69] X. Li and A. Scaglione, "Convergence and Applications of a Gossip-Based Gauss-Newton Algorithm", *IEEE Transactions on Signal Processing*, vol. 61, no. 21, pp. 5231-5246, 2013.

[70] L. Duc-Hung, P. Cong-Kha, N. Thi Thien Trang and B. Trong Tu, "Parameter extraction and optimization using Levenberg-Marquardt algorithm", in *2012 Fourth International Conference on Communications and Electronics (ICCE)*, 2012.

[71] F. Seco, A. R. Jimenez, C. Prieto, J. Roa and K. Koutsou, "A survey of mathematical methods for indoor localization", in *Proc. 2009 IEEE International Symposium on Intelligent Signal Processing*, 2009.

[72] H. Cox, "On the estimation of state variables and parameters for noisy dynamic systems", *IEEE Trans. Autom. Control*, vol. 9, no. 1, pp. 5-12, 1964.

[73] E. A. Wan and R. Van Der Merwe, "The unscented Kalman filter for nonlinear estimation", in *Proc. IEEE Adapt. Syst. Signal Process. Commun. Control Symp*, pp. 153-158, 2000.

[74] D. Simon, "A game theory approach to constrained minimax state estimation", *IEEE Transactions on Signal Processing*, vol. 54, no. 2, pp. 405-412, 2006.

[75] J. S. Liu and R. Chen, "Sequential montecarlo methods for dynamic systems", *J. Amer. Statist. Assoc.*, vol. 93, no. 443, pp. 1032-1044, 1998.

[76] D. Gualda, J. Ureña, J. García, and A. Lindo, "Locally-referenced ultrasonic-LPS for localization and navigation", *Sensors*, vol. 14, no. 11, pp. 21750-21769, 2014.

[77] A. Mahajan and F. Figueroa, "An automatic self-installation and calibration method for a 3D position sensing system using ultrasonics", *Robotics and Autonomous Systems*, vol. 28, no. 4, pp. 281-294, 1999.

[78] A. Nishitani, Y. Nishida, T. Hori and H. Mizoguchi, "Portable ultrasonic 3D tag system based on a quick calibration method", in *Proc. of the 2004 IEEE International Conference on Systems, Man. And Cybernetics*, pp. 1561-1568, 2004.

[79] J. Ureña, D. Ruiz, J. C. García, J. J. García, A. Hernández and M. C. Pérez, "LPS self-calibration method using a mobile robot", in *Proc. of the 2011 IEEE International Instrumentation and Measurement Technology Conference (I2MTC)*, pp. 1-6, 2011.

[80] D. Gualda, *Simultaneous Calibration of Ultrasonic Local Positioning Systems and Mobile Robot Navigation*, Ph.D. dissertation, Electronics Department, University of Alcalá, 2016.

[81] F. Zampella, A. Ramón, J. Ruiz and F. S. Granja, "Indoor Positioning Using Efficient Map Matching, RSS Measurements, and an Improved Motion Model", *IEEE Transactions on Vehicular Technology*, vol. 64, no. 4, pp. 1304-1317, 2015.

[82] J. Pinchin, C. Hide, and T. Moore, "A particle filter approach to indoor navigation using a foot mounted inertial navigation system and heuristic heading information", in *Proc.of the 2012 International Conference on Indoor Positioning and Indoor Navigation (IPIN)*, pp.13-15, 2012.

[83] D. Gualda, J. Ureña and E. García, "Partially constrained extended Kalman filter for navigation including mapping information", *IEEE Sensors Journal*, vol. 16, no. 24, pp. 9036-9046, 2016.

[84] D. Gualda, J. Urena, J. C. Garcia, J. Alcala and A. N. Miyadaira, "Reduction of ultrasonic indoor localization infrastructure based on the use of graph information", in *Proc. of 2016 International Conference on Indoor Positioning and Indoor Navigation (IPIN)*, 2016.

[85] D. Gualda, J. Ureña, J. C. Garcia, J. Alcala and A. N. Miyadaira, "Calibration of Beacons for Indoor Environments based on a Map-Matching Technique", in *Proc. of 2016 International Conference on Indoor Positioning and Indoor Navigation (IPIN)*, 2016.

[86] E. García, P. Poudereux, A. Hernández, J. J. García and J. Ureña. "DS-UWB indoor positioning system implementation based on FPGAs", *Sensors and Actuators A: Physical*, vol. 201, pp. 172-181, 2013.

[87] E. M. Gorostiza, J. L. Lázaro Galilea, F. J. Meca Meca, D. Salido Monzú, F. Espinosa Zapata and L. Pallarés Puerto, "Infrared sensor system for mobile-robot positioning in intelligent spaces", *Sensors*, vol. 11(5), pp. 5416-5438, 2011.

[88] Estefania Munoz Diaz, "Inertial Pocket Navigation System: Unaided 3D Positioning", *Sensors*, vol. 15, no. 4, pp. 9156-9178, 2015.

[89] Alfonso Bahillo, Teodoro Aguilera, Fernando J. Álvarez & Asier Perallos. "WAY: Seamless Positioning Using a Smart Device", Wireless Personal Communications, vol. 90, no. 2, pp. 1-19, 2016.

[90] F. Ijaz, Hee Kwon Yang, A. W. Ahmad and Chankil Lee, "Indoor positioning: A review of indoor ultrasonic positioning systems", in *2013 15th International Conference on Advanced Communication Technology (ICACT)*, pp. 1146-1150, 2013.

[91] J. Rabaey, "Reconfigurable processing: the solution to low-power programmable DSP", in *Proc. of the 22nd IEEE International Conference on Acoustics Speech and Signal Processing (ICASSP)*, pp. 275-278, 1997.

[92] S. Knapp, "Field configurable system on chip devices architecture", in *Proc. of the 22nd IEEE Custom Integrated Circuits Conference (CICC)*, pp. 155-158, 2000.

[93] J. Becker, N. Libeau, T. Pionteck and M. Glesner, "Efficient mapping of pre-synthesized IP-cores onto dynamically reconfigurable array architectures", in *Proc of the 11th International Conference on Field-Programmable Logic and Applicacions (FPL)*, pp. 584-589, 2001.

[94] Á. Hernández, E. García, D. Gualda, J. M. Villadangos, F. Nombela and J. Ureña, "FPGA-Based Architecture for Managing Ultrasonic Beacons in a Local Positioning System", *IEEE Trans. on Instrumentation and Measurement*, vol. 66, no. 8, pp. 1954-1964, 2017.

[95] Pro-Wave Electronics Corporation, *Air Ultrasonic Ceramic Transducers 328ST/R160*, Product Specification, 2005.

[96] E. García, J. A. Paredes, F. J. Álvarez, M. C. Pérez and J. J. García, "Spreading sequences in active sensing: A review", *Signal Processing*, vol. 106, pp. 88-105, 2015.

[97] GEINTRA-US/RF, Ultrasonic Indoor Positioning System, https://www.youtube.com/watch?v=d_ddYMlOp7g, 2009.

[98] GEINTRA-US/RF, Ultrasonic LPS using Generalized Cross-Correlation, https://www.youtube.com/watch?v=JO2yzZkLQK4, 2013.

[99] GEINTRA-US/RF, Ultrasonic positioning using GCC-PATH filter, https://www.youtube.com/watch?v=Wd20FHN7QM4, 2013.

[100] GEINTRA-US/RF, Multiple Ultrasonic Local Positioning Systems, https://www.youtube.com/watch?v=HVT9j5uEaq8, 2013.

[101] GEINTRA-US/RF, ALPS self-calibration, https://youtu.be/fstIf9jzKsM, 2016.

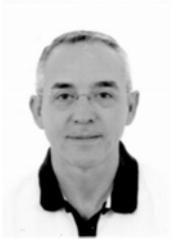
**Jesús Ureña (M'06, SM'15)** received the B.S. degree in Electronics Engineering and the M.S. degree in Telecommunications Engineering from Universidad Politécnica de Madrid, Spain, in 1986 and 1992, respectively; and the Ph.D. degree in Telecommunications from the Universidad de Alcalá, Spain, in 1998. Since 1986, he has been with the Department of Electronics, University of Alcalá, currently as a Professor. He has collaborated in several educational and research projects in the area of electronic control and sensorial systems for mobile robots and wheelchairs and in the area of electronic distributed systems for railways. His current research interests are in the areas of ultrasonic signal processing, local positioning systems (LPSs) and activity monitoring, and sensory systems for railway safety.

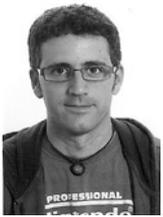
**Álvaro Hernández (M'06, SM'15)** received the Ph.D. degree from the University of Alcalá, Spain, and Blaise Pascal University, France, in 2003. He is currently an Associate Professor of Digital Systems and Electronic Design with the Electronics Department, University of Alcalá. His research areas are multisensor integration, electronic systems for mobile robots, and digital and embedded systems.

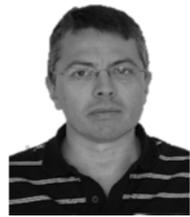
**J. Jesús García (M'06)** obtained his B.Sc. degree in Electronic Engineering from the University of Alcala (Spain) in 1992, and his M.Sc. degree from Polytechnic University of Valencia (Spain) in 1999. He received the Ph.D. degree with distinction from University of Alcala (Spain) in 2006. He is currently an Associate Professor of Digital and Analog Electronic at the Electronics Department of the University of Alcala. His research areas are multi-sensor integration, local positioning systems and biomedical signal processing.

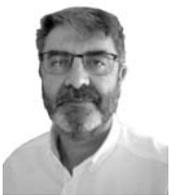
**José M. Villadangos** received the Ph.D. degree from the University of Alcala, Alcalá de Henares, Spain, in 2013. He is currently an Associate Professor of Electronic Digital Systems with the Electronics Department, University of Alcala. His current research interests include multisensor integration, electronic systems for mobile robots, digital and embedded systems, low-level ultrasonic signal processing, and local positioning systems.

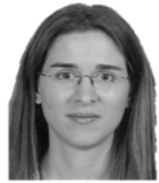
**Mª del Carmen Pérez (M'07)** received the M.S. degree in Electronics Engineering and the PhD degree from the University of Alcalá (UAH), Spain, in 2004 and 2009, respectively. She is currently an Associate Professor at the Electronics Department of the University of Alcalá. Since 2003 she has collaborated on several research projects in the areas of sequence design, low-level ultrasonic signal processing and computing architectures.

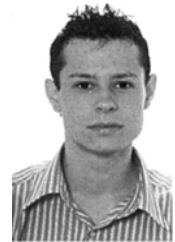
**David Gualda** received the B.S. degree in Electronics Systems and the M.Sc. degree in Advanced Electronics System, from the University of Alcala (Spain), in 2009 and 2011, respectively; and the Ph.D. degree in Electronics in 2016. He is currently at the Department of Electronics of the University of Alcalá with a Postdoctoral Research Contract. His main research interests are in the areas of ultrasonic indoor location, signal processing and information fusion

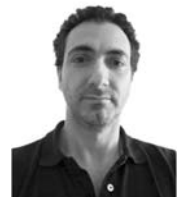
**Fernando J. Álvarez (M'07, SM'17)** received his Physics degree from the University of Sevilla (Spain), his Electronic Engineering degree from the University of Extremadura (Spain), his M.S. degree in Signal Theory and Communications from the University of Vigo (Spain), and his Ph.D. degree in Electronics from the University of Alcalá (Spain). During 2008 he was a postdoctoral 'José Castillejo' Fellow in the Intelligent Sensors Laboratory, Yale University (USA). He is currently an Associate Professor of Digital Electronics in the Department of Electrical Engineering, Electronics and Automation at the University of Extremadura, Spain, where he is also the head of the Sensory Systems Group. His research areas of interest include local positioning systems, acoustic signal processing and embedded computing.

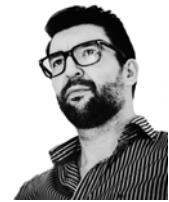
**Teodoro Aguilera** received his Physics degree in 2011 and his Ph.D. in Electronics in 2016 all of them from the University of Extremadura (Spain). He was an Assistant Professor of Automation for three years in the Department of Electrical Engineering, Electronics and Automation at the University of Extremadura. Currently he is a research member of the Sensory Systems Group of this university where his work is focused on the design of Acoustic Local Positioning Systems (ALPS) based on mobile devices.